\documentclass[epjCONF,columns]{svjour}
\usepackage[latin1]{inputenc}
\usepackage{color}
\usepackage{amsbsy}
\usepackage{amssymb}
\usepackage{graphicx}
\usepackage[numbers]{natbib}

\makeatletter

\providecommand{\tabularnewline}{\\}

 \journalname{epjCONF}
\usepackage{graphics}
\usepackage[varg]{txfonts}
\usepackage{jour-abbrev}
\session-title{Conference Title, to be filled}

\makeatother

\begin{document}
\abstract{The disruption of a star by the tidal field of a massive
  black hole is the final outcome of a chain of complex dynamical
  processes in the host galaxy. I introduce the ``loss cone problem'',
  and describe the many theoretical and numerical challenges on the
  path of solving it. I review various dynamical channels by which
  stars can be supplied to a massive black hole, and the relevant
  dynamical relaxation / randomization mechanisms. I briefly mention
  some ``exotic'' tidal disruption scenarios, and conclude by
  discussing new dynamical results that are changing our understanding
  of dynamics near a massive black hole, and may well be relevant for
  tidal disruption dynamics.\\ 
}

\title{Stellar dynamics and tidal disruption events in galactic nuclei}

\author{Tal Alexander\fnmsep%
\thanks{\email{tal.alexander@weizmann.ac.il}%
}}

\institute{Weizmann Institute of Science, Faculty of Physics, Department of
Particle Physics and Astrophysics}

\maketitle

\section{Outline}

\label{s:intro} This is a brief, informal review of the stellar dynamical
mechanisms that play a role in the tidal destruction (TD) of stars
by massive black holes (MBHs) in galactic nuclei. The key issue is
to determine how, and at what rate, are stars deflected to the extremely
eccentric orbits that bring them sufficiently close to the MBH for
TD. This is the so called ``loss-cone (LC) replenishment problem'',
which is introduced in its most basic form (single stars around a
single MBH, scattered singly into the LC) in Sec. ~\ref{s:LC}, together
with a discussion of the more general framework of infall / inspiral
processes, and other channels of LC replenishment. Orbital evolution
(denoted here generally by the term ``relaxation''), whether stochastic
or coherent, can take various forms and proceed at a wide range of
rates on different spatial scales. Some lesser known forms of relaxation
are presented in Sec. \ref{s:relax}, and their impact on TD is discussed.
TD rate estimates, and the rates of associated ``exotic'' processes,
are reviewed in Sec. \ref{s:TDrate}. Sec. \ref{s:tight} explores
some interesting new dynamical processes that may be relevant for
stars that are formed or captured very close to the MBH, and deflected
to the LC from tight orbits. Sec. \ref{s:summary} briefly summarizes
the main conclusions and offers a few final comments.

\section{Getting stars to the MBH: the loss cone}

\label{s:LC}

\subsection{Basic results}
\label{ss:basic}

TD occurs when a star approaches an MBH of mass $M_{\bullet}$ closer
than the TD radius $r_{t}\simeq
R_{\star}(M_{\bullet}/M_{\star})^{1/3}$, where $R_{\star}$ and
$M_{\star}$ are the star's radius and mass. Alternative physical
formulations of the TD criterion are that the star is destroyed when
its typical density, $\rho_\star \sim M_{\star}/R_{\star}^3$, falls below the density
the MBH would have had if its mass were spread over the volume
$r_t^3$, or that the star is destroyed when the crossing time through
the disruption zone $(r_t^3/GM_\bullet)^{1/2}$ falls below the star's
free-fall time $(R_\star^3/GM_\star)^{1/2}$.  TD by a MBH can occur
only when $r_{t}$ lies outside the event horizon at
$r_{e}=xGM_{\bullet}/c^{2}$, where $x=2$ for a non-spinning MBH, and
can be as small as $x=1$ for a maximally spinning MBH and a
co-orbiting star. 
Since $r_t/r_e\propto \rho_\star^{-1/3}x^{-1}M_\bullet^{-2/3}$,
it follows that the less massive the MBH, the
more compact the stars that undergo TD can be; for a star with given
$R_{\star},M_{\star}$, there exists a maximal MBH mass, $\max
M_{\bullet}$, where TD is still possible; and the higher the MBH spin,
the larger is $\max M_{\bullet}$.

\begin{figure*}
\begin{tabular}{ccc}
\raisebox{2em}{\includegraphics[width=0.31\textwidth]{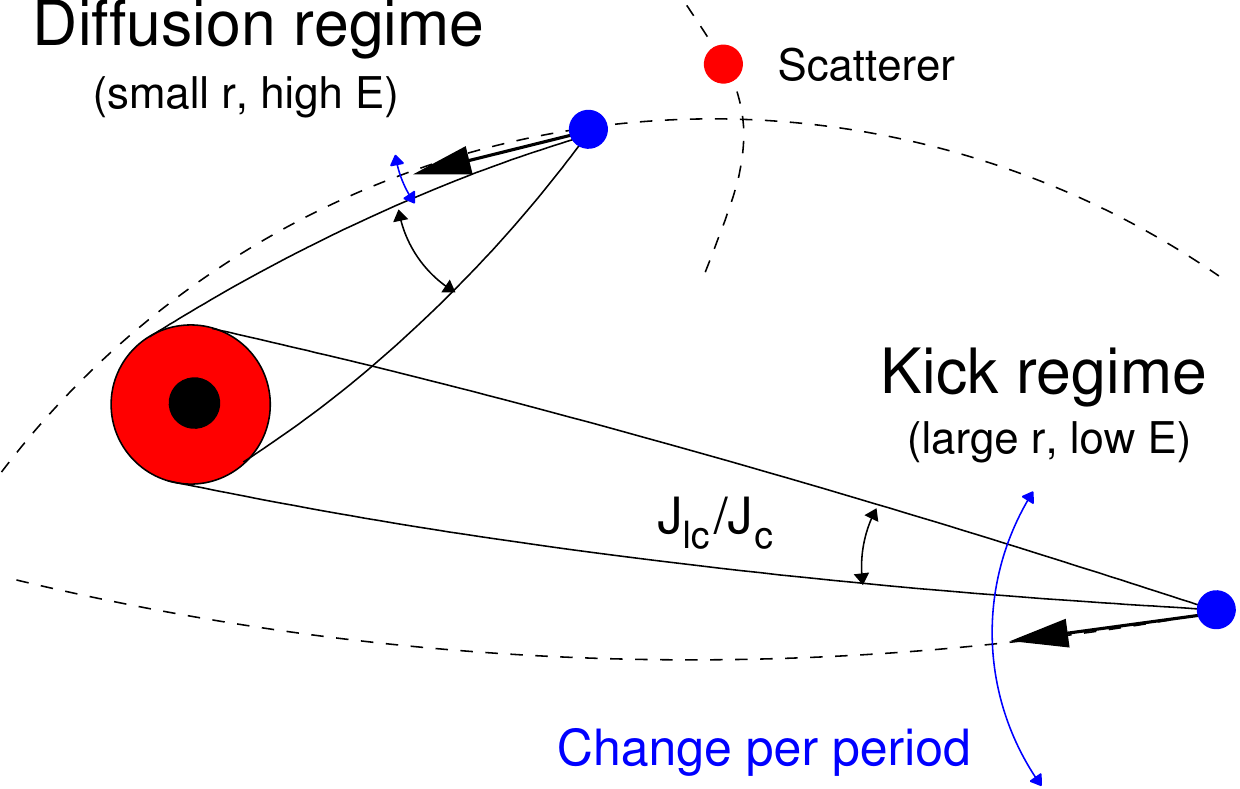}} & \includegraphics[width=0.31\textwidth]{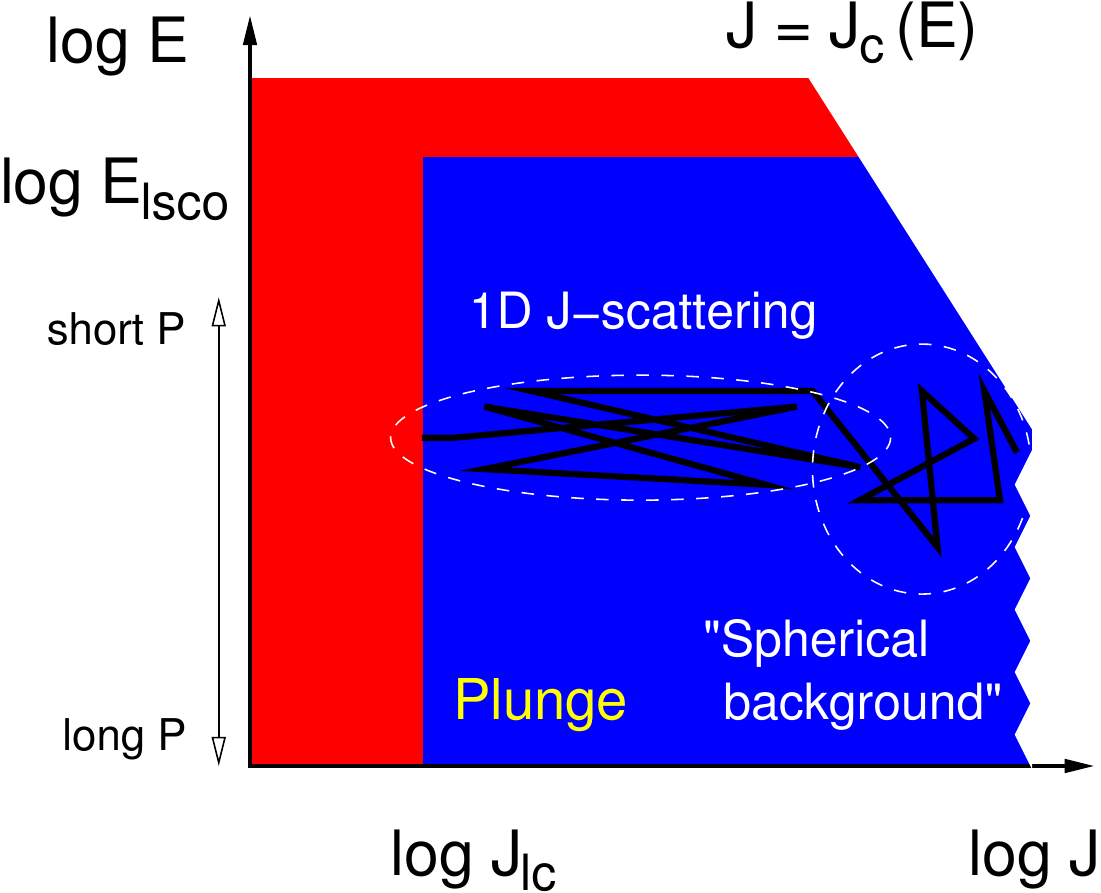} & \includegraphics[width=0.31\textwidth]{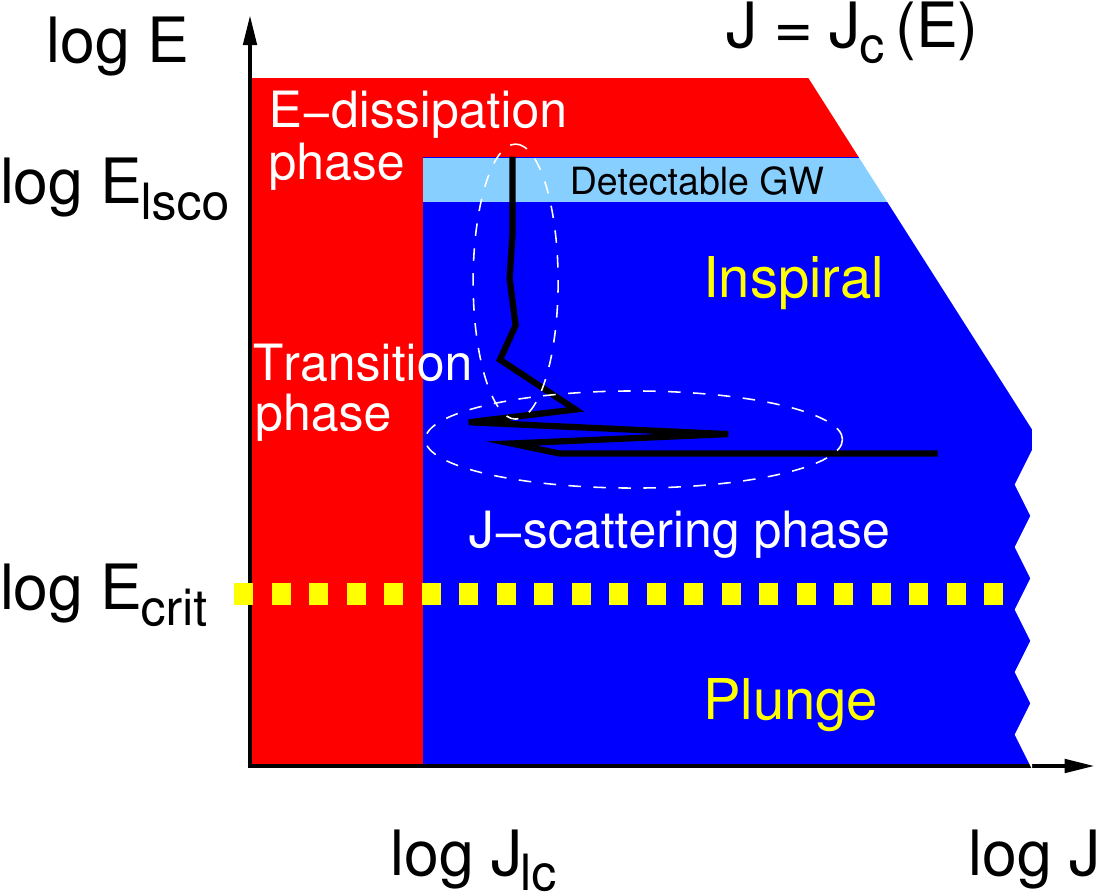}\tabularnewline
\end{tabular}

\caption{Schematic depiction of LC dynamics in real space and in phase
  space, and of the difference between infall and inspiral.\textbf{
    Left}: A test star's velocity vector is deflected into the LC by
  an interaction with a field star (scatterer). When the test star is
  relatively close to the MBH, the angular opening of the LC is large,
  while the magnitude of a typical accumulated deflection per orbit is
  small; stars diffuse into the LC at a rate much slower than the time
  it takes a star to reach the MBH, and the LC is almost empty of
  stars. Conversely, for stars far from the MBH, the angular size is
  smaller than the typical accumulated deflection per orbit, stars are
  rapidly ``kicked'' into the middle of the LC, which remains almost
  full. In typical galactic nuclei, most tidally disrupted stars
  originate from the transition range between the empty and full LC
  regimes, at $r\sim r_{h}$, where $N_{\star}M_{\star}\sim
  M_{\bullet}$. \textbf{Center}: The phase space $(\log J,\log E$) of
  infall (plunge) events, such as tidal disruption, which happen
  promptly once the star gets close enough to the MBH. Since
  $t_{J}\sim(J/J_{c})^{2}t_{E}$, once a star reaches a somewhat
  eccentric orbit, there is a substantial probability that it will be
  scattered into the LC and promptly destroyed. \textbf{Right}: The
  behavior in the presence of a dissipative process, for example the
  emission of gravitational waves (GW), whose luminosity sharply
  increases with proximity to the MBH. Close enough to the MBH, above
  some critical energy $E_{\mathrm{crit}}$, (equivalently, inside some
  critical radius $r_{\mathrm{crit}}$), where the dissipation per
  period is strong, orbital decay is faster that the 2-body
  relaxation, and almost all stars eventually inspiral gradually into
  the MBH. Conversely, outside that region, almost all stars
  ultimately plunge into the MBH. }

\label{f:LC} 
\end{figure*}

A star on an orbit with specific energy $E=GM_{\bullet}/2a$ and specific
angular momentum $J=J_{c}\sqrt{1-e^{2}}$ (The Keplerian limit is
assumed, with the convention that $E=-E_{\mathrm{true}}>0$ for a
bound orbit. $a$ is the semi-major axis (sma), $e$ the eccentricity,
and $J_{c}=\sqrt{GM_{\bullet}a}$ is the specific circular angular
momentum) will reach periapse inside the TD radius, $r_{p}=a(1-e)\le r_{t}$
if its angular momentum is less than the LC angular momentum $J_{lc}=\sqrt{GM_{\bullet}r_{t}(1+e)}$
(Fig. \ref{f:LC}L). All stars that are initially on LC orbits will
be destroyed in less than an orbital period $P$. From that time on,
the rate of TD events will depend on the rate at which dynamical mechanisms
repopulate these orbits. 

The discreteness of a stellar system guarantees a minimal orbital
randomization rate by incoherent 2-body relaxation, on the timescale
$t_{E}\sim Q^{2}P/N_{\star}\log Q$, where $Q=M_{\bullet}/M_{\star}$
and $N_{\star}(a)$ is the number of stars with sma $\le a$ (here and
below I informally interchange $r\leftrightarrow a$). This is the time
it takes the orbital energy to change by order unity. Because the
2-body scattering is local and isotropic (i.e. it does not ``care''
where the MBH is), the time for an order unity change in angular
momentum can be much shorter than $t_E$ if the orbit already has low
angular momentum, $t_{J}\sim[J/J_{c}(E)]^{2}t_{E}$. It then follows
that stars get deflected to LC orbits primarily by relaxation of
angular momentum, and not energy (Fig. \ref{f:LC}C).

More detailed calculations \citep[e.g.][]{Sye+99} indicate that the
total TD rate is $\Gamma_{\mathsf{TD}}\!\sim\!\left.N_{\star}(<r_{h})\right/\left\langle \log(J_{c}(a)/J_{lc})t_{E}(a)\right\rangle $,
where $r_{h}$ is the MBH's radius of influence, $r_{h}\sim GM_{\bullet}/\sigma^{2}$,
where $\sigma$ is the velocity dispersion far from the MBH, and where
the average $\left\langle \cdots\right\rangle $ is over the volume
within $r_{h}$. Note that the dependence of the rate on the relative
size of the LC is only logarithmic, so that the TD rate scales approximately
as the number of stars in the relevant reservoir, over the relaxation
time on that scale.

\subsection{Plunge vs Inspiral}

\label{ss:P_vs_I}

TD is one example of an \emph{infall} process, where the event of
interest occurs promptly once the star reaches close enough to the
MBH. It is instructive to consider also the related class of
\emph{inspiral }processes, where energy is dissipated gradually on
successive peripassages, and the orbit decays until the star is
destroyed. The dissipation can be, for example, by tidal heating
(Sec. \ref{ss:exotica}), or by the emission of gravitational waves
(GW). In both cases the dissipated power is a strong function of the
distance from the MBH, and therefore almost all of it occurs near
periapse. Since inspiral is a race between deterministic dissipation
and stochastic scattering, there exists a boundary, which is
approximately constant in energy (equivalently, a typical radius from
the MBH), that separates infall from inspiral (Fig.\ref{f:LC}R). Far
from the MBH, where the orbital time is long, there is enough time
between peripassages to be scattered off the nearly radial orbit
required for a close encounter with the MBH. Therefore stars beyond
some critical radius $r_{\mathrm{crit}}$ ultimately fall directly
(plunge) into the MBH. Conversely, stars inside $r_{\mathrm{crit}}$
inspiral gradually into the MBH, at a rate
$\Gamma_{\mathrm{inspiral}}\sim N_{\star}\left[<\!
  r_{\mathsf{crit}}(t_{E})\right]\left/\left\langle
\log(J_{c}/J_{lc})t_{E}\right\rangle \right.$. This expression, whose
interpretation is similar to that of the infall rate, is deceptively
simple, since it is non-linear because $r_{\mathrm{crit}}$ depends
itself on $t_{E}$. For GW inspiral in the GC, $r_{\mathrm{crit}}\sim
O(0.01\,\mathrm{pc})$ (Fig. \ref{f:MP}R). Because typically
$r_{\mathrm{crit}}\ll r_{h}$, and the number of stars enclosed within
$r_{\mathrm{crit}}$ is much smaller than in $r_{h}$, it is generally
the case that $ $ $\Gamma_{\mathsf{inspiral}}\sim
O(0.01)\Gamma_{\mathsf{plunge}}$.

\subsection{The challenges}

\label{s:challenge}Many conceptual, technical, and astrophysical
challenges stand in the way of reliable theoretical predictions of
TD rates and their parameter dependences. These issues and open questions
broadly fall in three categories:

\paragraph{Analytic representation }

How to achieve a tractable self-consistent description of the stellar
system? How to set boundary conditions correctly? How to describe
the diffusion of orbital elements in phase space?

\paragraph{Numeric simulation}

How to simulate the $N_{\star}\to\infty$ limit? How to deal with
the $r\to0$ limit (very close encounters; General Relativistic (GR)
effects)?

\paragraph{Galactic models}

Which are the dominant components (e.g. one or two MBHs; gas; disks)?
What stellar population to assume (mass spectrum)? What stellar density
to assume (distribution function)? What symmetries to assume (e.g.
spherical, axial)?\\

While dealing with questions of analytic representation and numeric
simulation may require sophisticated analysis and state of the art
computational techniques, these issues are in principle solvable, even
if only by brute force computation. In contrast, the situation with
the astrophysical uncertainties of galactic modeling presents an
irreducible problem, since it is unlikely that these could be
determined with any degree of certainty in the foreseeable future. At
this time there is no recourse but to explore a large parameter space
of plausible scenarios, and map out the range of possible outcomes.

\subsection{Supply channels}

Although much of the theoretical effort of modeling TD events focused
on the simplest configuration: stars that are singly deflected toward
a single MBH%
\footnote{This dynamical configuration also applies to stars and white dwarfs
  disrupted by intermediate mass black holes (IMBHs), or comets
  disrupted by a central object.  } (Sec. \ref{ss:1MBH}), there are
quite a few additional possible channels leading to TD events. These
include binary stars separated by the tidal field of an MBH, or
interactions of stars with a circum-nuclear gas disk or star disk
around a MBH. All of the above can also be relevant when the MBH is
not stationary in the center of the galaxy, but is recoiling following
an asymmetric coalescence event, or in the case where the stars
surround a binary MBH (Sec. \ref{ss:2MBH}). Many of these supply
channels are discussed in detail elsewhere in these proceedings.

\section{Rapid relaxation mechanisms}

\label{s:relax}

TD rates are determined by the supply channel, and by the replenishment
rate of the LC. 2-body relaxation is unavoidable, and thus sets a
guaranteed minimal replenishment rate. However, other processes can
contribute, or even dominate relaxation. Much of the motivation for
the study of these processes, and most of the empirical tests of these
ideas, are due to the unique observations of stars in the Galactic
Center (GC), very close to the Galactic MBH \linebreak{}
(SgrA$^{\star}$) \citep{ale05,gen+10} (Fig. \ref{f:MP}L). I briefly
describe here rapid relaxation by massive perturbers (MPs), by resonant
relaxation (RR), and orbital evolution in non-spherical systems.

\subsection{Massive perturbers}

\label{ss:MPs}

the rate of relaxation of a star of mass $M_{\star}$ by 2-body
encounters with perturbers of number density $n_{p}$ and mass $M_{p}$
can be estimated by a simple ``$\Gamma\sim nv\Sigma$'' collision rate
calculation. Approximating the relative velocity $v$ by the velocity
dispersion $\sigma$, and the interaction cross-section $\Sigma=\pi
R^{2}$ with the collision radius $R\sim GM_{p}/\sigma^{2}$ (the
closest approach needed for an $O(1\,\mathrm{rad})$ angular
deflection) yields a relaxation rate of $\Gamma\propto
n_{p}M_{p}^{2}/\sigma^{3}$ (integration over collisions at larger
separations increases the rate by the Coulomb factor). The $M_{p}^{2}$
dependence of the rate implies that even a small density of very
massive objects can dominate 2-body relaxation by stars, when
$n_{p}M_{p}^{2}/n_{\star}M_{\star}^{2}\gg1$ \citep{per+07}. This is
indeed the case in many galactic nuclei, where giant molecular clouds
(GMCs), stellar clusters (and possibly IMBHs) lower the estimated
2-body relaxation time by orders of magnitude relative to that due to
stars alone.

\begin{figure*}
\noindent \begin{centering}
\begin{tabular}{ccc}
\includegraphics[width=0.3\textwidth]{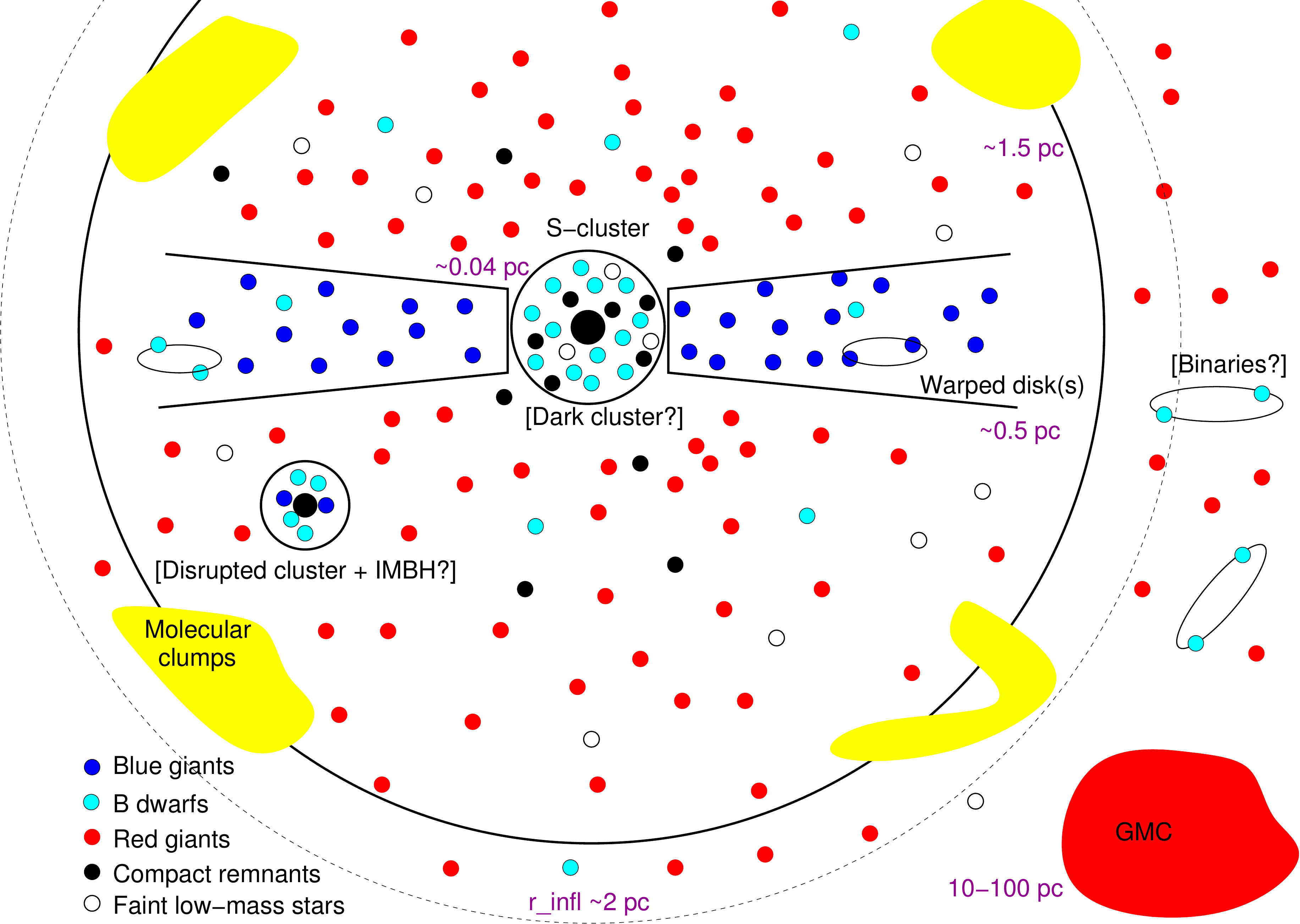} & \includegraphics[width=0.3\textwidth]{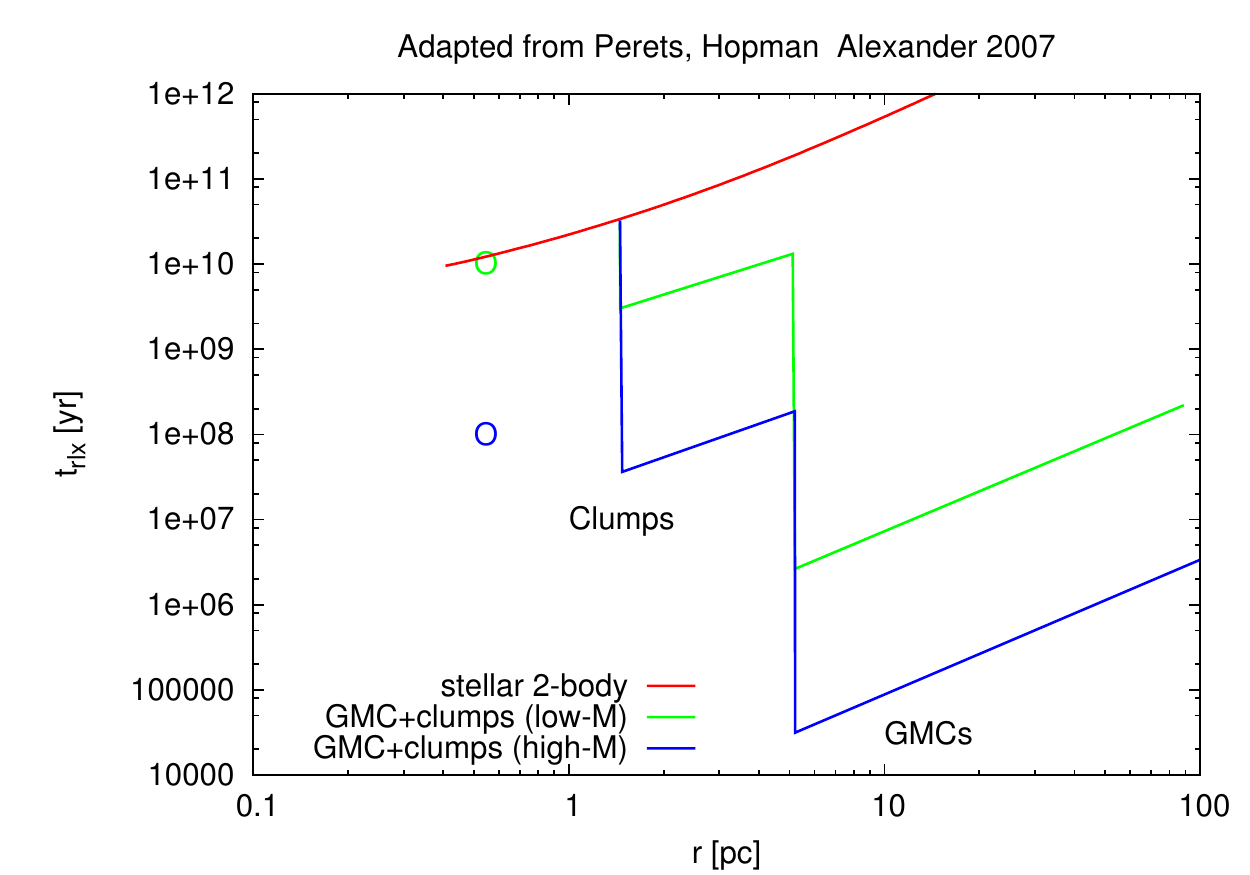} & \includegraphics[width=0.3\textwidth]{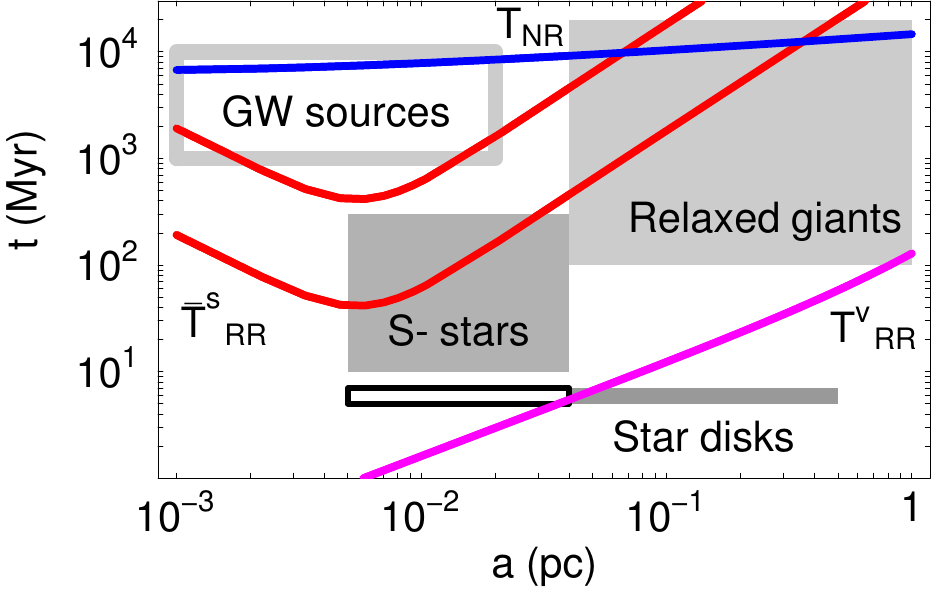}\tabularnewline
\end{tabular}
\par\end{centering}

\caption{\label{f:MP}Lessons from the GC. \textbf{Left}: A
  not-to-scale schematic of the various dynamical components that are
  observed in the GC around SgrA$^{\star}$, or are hypothesized to
  exist there (marked by {[}$\ldots${]}) \citep{ale11}. Note in
  particular the massive GMCs and the circum-nuclear molecular disk at
  or outside the MBH's radius of influence; the mixed old and young
  stellar population inside $r_{h}$ (only the giants and B-dwarfs are
  above the present-day detection threshold); the ${\cal O}(100)$ blue
  giants that are concentrated in one (or perhaps two)
  coherently-rotating warped disks, which extend inward to a sharp
  inner cut-off at $\sim0.04$ pc ($\sim1"$); and the spectroscopically
  and dynamically distinct population of $\sim40$ B-dwarfs on
  isotropic orbits inside $\sim1"$ (the ``S-cluster'').\textbf{
    Center}: The GMCs and molecular clumps of the circum-nuclear disk
  dramatically shorten the relaxation time outside $r_{h}$ (blue and
  green lines, for two limiting assumptions about the masses of the
  GMCs), relative to that by stars alone (red line), and also reduce
  from afar the relaxation time inside $r_{h}$ (circles)
  \citep{per+07,ale11}. \textbf{Right}: Possible signature of RR in
  the inner pc of the GC \citep{hop+06a}. Significant relaxation is
  expected where the relaxation times are shorter than the typical
  ages of the different dynamical component. 2-body relaxation is
  faster than scalar ($J$-changing) RR only far from the MBH, where
  the enclosed stellar mass becomes substantial, or extremely close to
  the MBH, where GR precession is strong. The two lines for scalar RR
  represent different assumptions about the mean stellar mass. Scalar
  RR is an important process for the dynamics of GW sources, and may
  be also implicated in the randomization of eccentricities of the
  S-stars \citep{per+09,bar+12b} and the relaxed giants. Vector
  ($\hat{\boldsymbol{J}}$-changing) RR is everywhere faster than
  2-body relaxation, as long as the average potential is close to
  spherical, Vector RR may explain the inner cutoff of the young star
  disk, and the randomization of the orbital inclinations of the
  relaxed giants. (reproduced with permission from the Astrophysical
  Journal).}
\end{figure*}

The acceleration of the relaxation rate in the GC, estimated from
observations of GMCs, is dramatic outside $r_{h}$, where GMCs are
found, and substantial also inside $r_{h}$ (Fig. \ref{f:MP}C). Since
on the $r\sim r_{h}$ scale, where most of the TD flux originates
(Sec. \ref{s:LC}; Fig. \ref{f:LC}L), the LC can already be filled by
stellar 2-body relaxation alone, MPs raise the TD rate only by $\times3$ over
that expected by stars alone (the effect is mostly due to the less
massive gas clumps at $\sim r_{h}$.  This conclusion could overturn if
IMBHs do exist inside $r_{h}$).

The most dramatic effect of MPs is on the scattering of binaries from
outside $r_{h}$ to LC orbits that lead to their tidal separation
(not to be confused with star-destroying TD), resulting in the capture
of one of the stars on a tight eccentric orbit around the MBH, and
the ejection of the other out of the Galaxy as a hyper-velocity star
\citep{bro+05}. A comparison of the numbers of bound B-dwarfs observed
in the inner $1^{"}$ of the GC (the ``S-stars'', see Fig. \ref{f:MP}L),
and the numbers of hyper-velocity stars (also mostly B-dwarfs), supports
this scenario for S-star capture by MP scattering \citep{per+07}.
While the capture process itself is ``dark'', the captured stars
are so close to the MBH, that a substantial fraction of them are likely
to eventually undergo luminous TD \citep{per+09,bar+12b} (Sec. \ref{s:tight}).

\subsection{Resonant relaxation}

\label{ss:RR}

RR is a rapid process of angular momentum relaxation \citep{rau+96,hop+06a},
which takes place in near-spherically symmetric potentials where orbital
evolution is limited (e.g. fixed ellipses in a Kepler potential; fixed
orbital planes in a general spherical potential). In that case, on
timescales much longer than $P$, but still shorter than $t_{\mathrm{coh}}$,
the time it takes for perturbations to grow and break the coherence
of the symmetry (e.g. deviations from a Kepler potential due to the
mass of the stars themselves, or due to GR precession), the averaged
orbits of the field stars can be viewed as fixed mass wires (in a
Kepler potential), or fixed annuli (time-averaged rosettes in a spherical
potential). 

The potential of this near-stationary background mass distribution
conserves orbital energy, but the finite number of stars means that
there are order $\sqrt{N_{\star}}$ fluctuations from isotropy, which
give rise to near-constant residual torques that rapidly ($\propto t$)
change the angular momentum of a test star. The accumulated change
over a coherence time, $(\Delta J)_{\mathrm{coh}}\sim(\sqrt{N_{\star}}GM_{\star}/R)t_{\mathrm{coh}}$,
where $R$ is the typical size of the test star orbit, then becomes
the mean free step size of an accelerated random walk, which can be
written as $(\Delta J/J_{c})(t)=\sqrt{t/t_{RR}}$ (for $t>t_{\mathrm{coh}}$),
where the RR timescale is $t_{RR}\sim Q^{2}P^{2}/(N_{\star}t_{\mathrm{coh}})$. 

The evolution of the angular momentum then depends on the symmetries
of the potential and the process that sets the shortest coherence
time. In a near-Kepler potential, all components of the angular
momentum are torqued, and in particular RR can change the magnitude
$J$, and drive a wide orbit into a near radial one that lies inside
the LC.  The relevant coherence-limiting (quenching) mechanisms are
precession due to the enclosed stellar mass far from the MBH, and GR
in-plane precession near it (Fig. \ref{f:MP}R). This variant of RR is
sometimes called ``scalar RR''. In addition, any spherical potential
supports a restricted form of RR, where the time-averaged mass annuli
change only the inclination of the orbital plane of a test star,
$\hat{\boldsymbol{J}}=\boldsymbol{J}/J$, but not the magnitude $J$. In
the absence of axial symmetry breaking, this is an extremely fast
process, since it is only quenched by the orbital randomization due to
RR itself\footnote{The separation of background stars / test star is
  arbitrary--all the stars in the system are simultaneously torqued by
  RR.  }. The self-quenching timescale $t_{\mathrm{coh}}^{sq}$ is then
set by the requirement $(\Delta J)_{\mathrm{coh}}\sim J_{c}$, which is
a long timescale, $t_{\mathrm{coh}}^{sq}\sim QP/\sqrt{N_{\star}}$.
This variant of RR is sometimes called ``vector RR''.

RR is easily detected in $N$-body simulations \citep{eil+09}, and
there is also some evidence that it is operating in the GC \citep{hop+06a}
(Fig. \ref{f:MP}R). Its direct effect on the integrated TD rates
is not large (up to $\times2$ enhancement \citep{rau+96}), since
most of the TD flux is from $r\sim r_{h}$, where $N_{\star}M_{\star}\sim M_{\bullet}$
and orbital precession due the enclosed stellar mass quenches RR.
However, RR may play a more significant role for TD events by stars
on tight orbits (Sec. \ref{s:tight}).

\subsection{Non-spherical systems}

\label{ss:nonSphere}

A high level of symmetry in the stellar potential is associated with
conserved quantities, which restrict orbital evolution and can slow
down LC refilling to the minimal rate provided by 2-body relaxation (a
collisional effect outside the domain of potential
theory). Conversely, less symmetry leaves more freedom for orbits to
evolve, and opens new possibilities for efficient replenishment of
loss-cone orbits.

Axisymmetric systems have centrophilic (center appro\-aching) orbits,
which lead to a $\times2$ enhancement in the TD rate over that
predicted for spherical systems \citep{mag+99}. Another consequence is
the higher incidence of deep TDs, with very small periapse distance,
which may result in tidal detonation (Sec. \ref{ss:exotica}). Triaxial
systems can support a large fraction pf chaotic orbits, which can lead
to a dramatic $\times10-100$ enhancement while the chaotic orbits
persist \citep{mer+04b}. This conclusion depends however on whether
chaotic orbits can exist in the presence of the stabilizing influence
of the central MBH.

\section{Tidal disruption rates}

\label{s:TDrate}

\subsection{Single MBH in steady state}

\label{ss:1MBH}

The basic configuration to consider is TD on a single MBH surrounded
by stellar cluster in steady state. TD rate estimates for such a configuration
have a long history, starting in the mid 1970's. Early analyses focused
on the similar process of a star disrupted by an IMBH in a globular
cluster \citep{fra+76,coh+78}. The first estimate for TDs on a MBH
in a galactic nucleus yielded a very high rate , of $10^{-3}-10\, M_{\odot}\,\mathrm{yr^{-1}\, gal^{-1}}$
\citep{You+77}. Contemporary rate estimates, with more careful modeling
of the process, yield consistently lower values. These are summarized
in table \ref{t:TDrate}.

\begin{table*}
\caption{\label{t:TDrate}Contemporary TD rate estimates for a single MBH in
steady state}

\noindent \begin{centering}
\begin{tabular}{lclc}
\hline 
\noalign{\vskip0.2em}
Special physical features & Rate {[}$ $$\mathrm{yr^{-1}}\,\mathrm{gal^{-1}}${]} & Comments & References\tabularnewline[0.2em]
\hline 
\noalign{\vskip0.2em}
\noalign{\vskip0.2em}
--- & $10^{-6}-10^{-4}$ &  & \citep{Sye+99}\tabularnewline[0.2em]
\noalign{\vskip0.2em}
\noalign{\vskip0.2em}
Axiality & $10^{-9}-10^{-4}$ &  & \citep{mag+99}\tabularnewline[0.2em]
\noalign{\vskip0.2em}
\noalign{\vskip0.2em}
Triaxiality / Chaos & $\lesssim10^{-2}$ &  & \citep{mer+04b}\tabularnewline[0.2em]
\noalign{\vskip0.2em}
\noalign{\vskip0.2em}
Resonant relaxation & --- & $\times2$ enhancement only & \citep{rau+96,rau+98}\tabularnewline[0.2em]
\noalign{\vskip0.2em}
\noalign{\vskip0.2em}
Revised $M_{\bullet}/\sigma$ relation & --- & $\times10$ increase over previous estimates & \citep{wan+04}\tabularnewline[0.2em]
\noalign{\vskip0.2em}
\noalign{\vskip0.2em}
Infall / inspiral & $9\times10^{-5}$ & Rate estimate for the GC & \citep{ale+03b}\tabularnewline[0.2em]
\noalign{\vskip0.2em}
\noalign{\vskip0.2em}
$N$-body experiments & $10^{-6}-10^{-4}$ &  & \citep{bro+11}\tabularnewline[0.2em]
\noalign{\vskip0.2em}
\noalign{\vskip0.2em}
MBH spin & --- & Higher TD rate by increased $\max M_{\bullet}$ & \citep{kob+04,iva+06,kes12}\tabularnewline[0.2em]
\hline 
\noalign{\vskip0.2em}
\end{tabular}
\par\end{centering}

\end{table*}

\subsection{Binary MBHs beyond steady state}

\label{ss:2MBH}

The near ubiquity of nuclear MBHs from very early cosmic times, and
the bottom-up structure formation paradigm, together imply that
MBH pairs and bound MBH binaries must be a commonly occurring, probably
transient phase in galaxy evolution. Recent studies have found that
hard, high mass ratio binary MBHs are efficient disruptors. This can
come about by Kozai forcing of stellar orbits around the primary (more
massive) MBH, induced by the secondary MBH \citep{iva+05}, or by
chaotic mixing due to the secondary \citep{che+11}. These effects
can lead to a transient phase of very high TD rates, $\lesssim1\,\mathrm{yr^{-1}}$,
and account overall for $O(10\%)$ of the steady state cosmic TD rate.
Another consequence of this high TD rate is the possibility of multiple
events from the same galaxy over an observationally relevant short
timescale, which could provide a signature of a binary MBH \citep{weg+11}.

Certain combinations of orbital angular momentum and MBH spin
orientations lead to a strong asymmetry in the emitted GW flux and a
very fast recoil velocity for the newly coalesced MBH \citep{sch+07}.
The recoiling MBH retains a small cluster of tightly bound stars with
it. The recoil can promptly refill the LC and briefly raise the TD
rate to $O(0.1\,\mathrm{yr^{-1}})$, thereby providing a prompt EM
counterpart to the GW signal of the coalescence \citep{sto+11}. Much
later, as the recoiling MBH moves farther away from the center of the
galaxy, the small dense cluster it carries with it can provide
off-center or even extra-galactic TD events at rates between $0.01$
\citep{sto+12} to $\lesssim1$ \citep{kom+08} of those of non-recoiling
MBHs.

\subsection{TD-related exotica}

\label{ss:exotica}

It is interesting to note that there are exotic processes that must
occur, and at substantial rates, in association with TD. These happen
when the orbital periapse is either just outside the tidal disruption
radius $r_{t}$, or very deep inside it.

\paragraph{Near misses}

Stars that are deflected from a tight enough orbit to an eccentric
one with periapse within $\sim2r_{t}$ can enter an inspiral phase,
similar to that of GW inspiral events, as ``squeezars'' \citep{ale+03a},
stars that brightly shine by the orbital energy dissipated as they
tidally interact with the MBH. This inspiral-type mechanism has a
rate that is typically $\sim0.05$ of the TD rate, $0.1-1\,\mathrm{yr^{-1}}$
in the GC (Sec. \ref{ss:P_vs_I}). When stars are not captured on
an inspiral orbit, they undergo a single very strong tidal scattering
interaction \citep{ale+01b}, which can spin them up and mix them,
thereby affecting their subsequent spectroscopic signature and evolution.
Such event are predicted to occur at $O(1)$ of TD rate, $10^{4}-10^{5}\,\mathrm{yr}^{-1}$
in GC. Finally, the extended, low-gravity envelopes of giant stars
are susceptible to tidal stripping, leaving behind the hot bare cores
\citep{dav+05}. These events are estimated to occur at $\sim0.01$
of the TD rate in the GC.

\paragraph{Direct hits}

When the periapse of the incoming star is $\ll r_{t}$, and the stellar
core is in a state that is favorable to fast nuclear burning processes
(post-hydrogen burning), the strong tidal pancake compression the
star experiences as it crosses periapse can lead to a supernova-like
tidal detonation event \citep{car+82,lag+93}. The rate of such events
is estimated at $<0.1$ of the TD rate.

\section{Tidal disruptions from tight orbits}

\label{s:tight}

Standard incoherent 2-body relaxation results in loss-cone refilling
that is dominated by stars from typical radii $r\sim r_{h}$, where
the dynamical timescale is long, the MBH potential no longer dominates
(and therefore RR is inefficient), and GR effects are completely negligible.
However, there are compelling reasons to believe that there are situations
where stars on much tighter orbits can contribute significantly to
the total TD rate. This can be the case, for example, when there is
substantial in-situ star formation in gravitationally unstable accretion
disks, as is observed in the GC on the $O(0.1r_{h})$ scale \citep{pau+06};
or for stars captured around the MBH by the tidal separation of binaries
\citep{hil88} (Sec. \ref{ss:MPs}); or for stars that are tightly
bound to a recoiling MBH \citep{kom+08,sto+12}.

In such cases, dynamical mechanisms that have not been considered before
may come into play. I focus here on two newly uncovered effects: stellar
relaxation by anomalous diffusion \citep{bar+12}, and the Schwarzschild
Barrier (SB) in phase space \citep{mer+11}. While it is too early
to tell whether these effects have a significant impact on TD events,
their recent discovery highlights the possibility that basic key components
may still be missing from our understanding of dynamics near MBHs. 

\begin{figure*}
\noindent \begin{centering}
\begin{tabular}{ccc}
\raisebox{1em}{\includegraphics[width=0.32\textwidth]{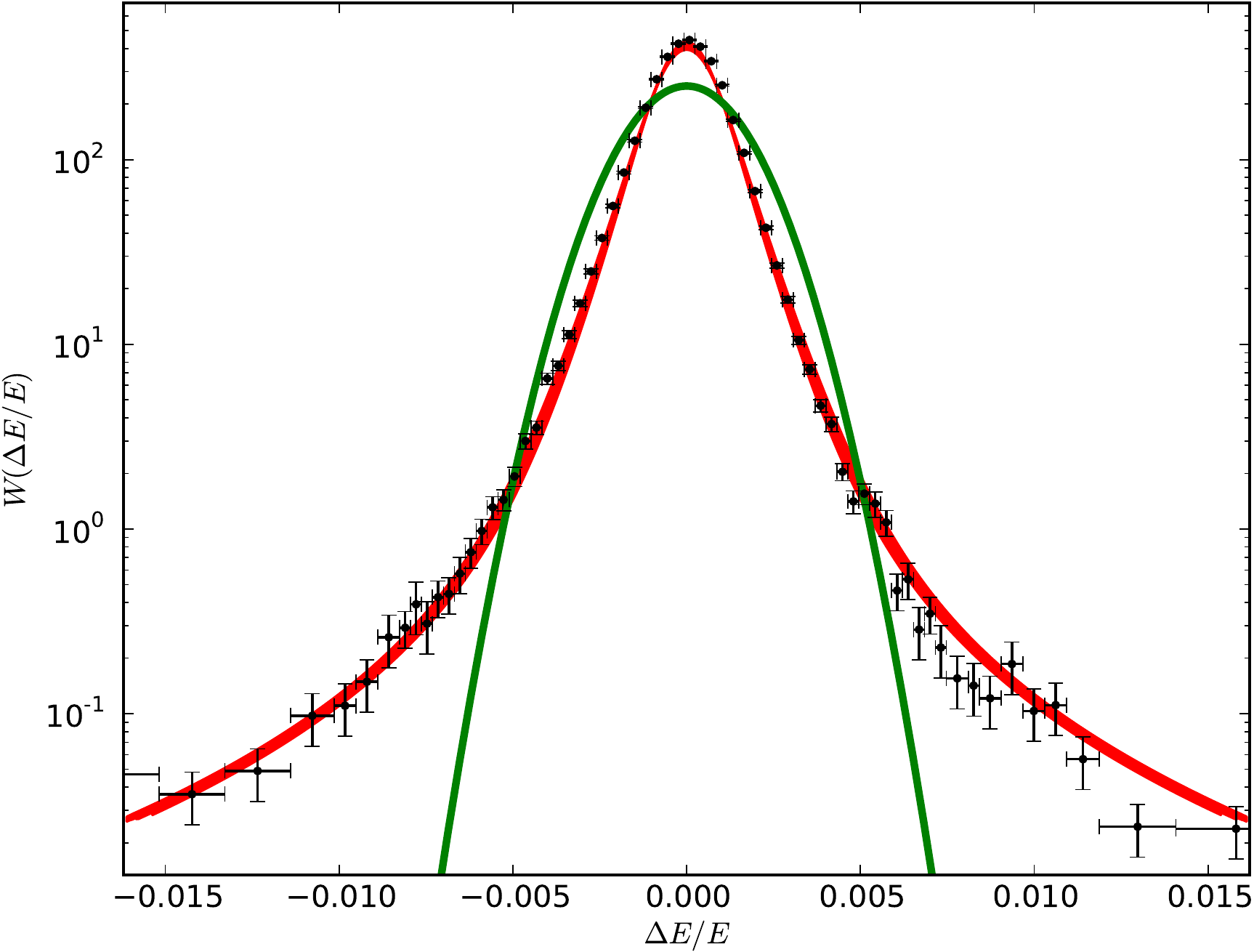}} & \includegraphics[width=0.31\textwidth]{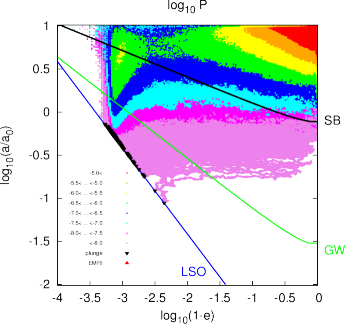} & \includegraphics[width=0.31\textwidth]{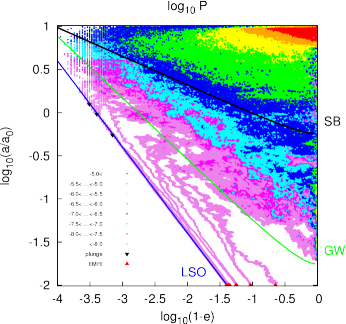}\tabularnewline
\end{tabular}
\par\end{centering}

\caption{\label{f:neweffects}Previously unconsidered dynamical effects
  that could modify TD rates from orbits near the MBH. \textbf{Left}:
  Anomalous diffusion on timescales shorter than $t_{E}$ leads to
  faster than $\propto\sqrt{t}$ evolution in phase space (crosses from
  $N$-body simulation and theory in red trace the relative change in
  energy $\Delta E/E$), and in particular makes ``high-$\sigma$''
  events much more common than they would be in normal diffusion
  (Gaussian probability distribution function in green)
  \citep{bar+12}.\textbf{ Center}: The phase-density of $50\,
  M_{\odot}$ stellar mass ``black holes'' around a
  $M_{\bullet}=10^{6}\, M_{\odot}$ MBH in $(\log[1-e],\log
  a/1\,\mathrm{mpc})\sim\mbox{(}\log J,\log E)$ phase space, as
  obtained by Monte Carlo experiments \citep{ale+12}. In the absence
  of GR, RR rapidly throws all stars into the MBH across the last
  stable orbit (LSO). \textbf{Right}: With GR in-plane precession, the
  Schwarzschild Barrier (a resonance between the RR torques and the GR
  precession) \citep{mer+11} deflects stars back to lower eccentricity
  orbits and allows gradual GW inspiral to take place (see tracks with
  nearly constant $r_p$ parallel to the LSO line). }
\end{figure*}

\paragraph{Anomalous diffusion}

Normal diffusion can be viewed as a random walk process, where the
accumulated change $\Delta E$ or $\Delta J$ grows with time as
$\propto\sqrt{t}$ (I omit here for brevity discussion of drift). This
is a consequence the Central Limit Theorem (CLT) applied to multiple
scattering events.  The CLT is valid since each of those is physically
limited to some \emph{finite} range of $\Delta E$ or $\Delta J$, and
is thus well-defined statistically. However, the \emph{rate} of
convergence to the CLT can be very slow, depending on how
nearly-di\-vergent is the physics of the scattering. In that case, on
timescales shorter than CLT convergence, the evolution of $E$ or $J$
is by anomalous diffusion. The rate can be very different from
$\sqrt{t}$; it may depend on $\Delta E$ itself (non-self-similar
evolution); it can result in improbably high rates of
``high-$\sigma$'' events (when mis-interpreted as Normal probability
events). The Newtonian gravitational force with its formal divergence
at $r_{12}\to0$ is exactly such a special case
(Fig. \ref{f:neweffects}L). On timescales shorter than $\sim0.1t_{E}$
there are substantial deviations from the expected statistics of
normal diffusion.

Anomalous diffusion may be particularly relevant close to the MBH,
where it could change the branching ratios between infall and inspiral
(effectively changing $r_{\mathrm{crit}},$ see
Sec. \ref{ss:P_vs_I}). This can be seen by noting that inspiral can
occur only when the dissipation timescale is shorter than the
relaxation time, $t_{\mathrm{diss}}\sim(E/\Delta
E_{\mathrm{diss}})P<t_{J}(J_{lc})$, where $\Delta E_{\mathrm{diss}}$
is the energy dissipated per peripassage, and where both $\Delta
E_{\mathrm{diss}}$ and $t_J$ are evaluated close to the LC boundary.

Since $t_{\mathrm{diss}}/t_{J}(J_{lc})\propto N_{\star} \propto
a^{3-\alpha}$ for an $n_{\star}\propto r^{-\alpha}$ cusp
(Sec. \ref{ss:basic}), then if anomalous diffusion in $\Delta J$ (not
yet explored) leads to overall faster $J$-relaxation, it will shift
$r_{\rm crit}$ to a smaller radius, implying more prompt TDs and fewer
tidal inspirals than estimated for normal diffusion.


\paragraph{The Schwarzschild Barrier}

RR torquing of orbits (Sec. \ref{ss:RR}; Fig \ref{f:MP}R) becomes
more important closer to the MBH, until it is quenched by GR in-plane
precession very close to it. It was hypothesized that this quenching
is crucial for ``saving'' stars from plunging into the MBH, and
allowing them to inspiral gradually (Fig. \ref{f:neweffects}C). Relativistic
$N$-body simulations \citep{mer+11} revealed that the GR precession
not only quenches RR and enables inspiral, but in fact creates a reflective
barrier in phase space, which can be approximately understood as arising
from a resonance between the GR precession frequency and the rate
of angular momentum change due to the RR torques (Fig. \ref{f:neweffects}R).
Stars can cross the SB only by 2-body scattering. Once past the barrier,
they are free to inspiral. Current analysis of the SB phenomenon has
focused only of GW inspiral. The implications for TD events from tight
orbits are still unknown.

\section{Summary}

\label{s:summary}

It is clear that there are multiple plausible channels by which stars
can be deflected into LC orbits that lead to TD events. However, quantitative
estimate of the cosmic rates of such events are severely limited by
astrophysical uncertainties. On the theoretical side, there is a clear
need for a better understanding of the conventional dynamical mechanisms
that are known to play part in the replenishment of the LC, and there
are also new, recently discovered mechanisms that need to be taken
into account. In spite of these uncertainties, it seems quite realistic
to expect that TD rates should be high enough to be observationally
interesting. Indeed there are multiple efforts to approach the problem
empirically, and detect TD events (see in these proceedings). 

As is evident from this review, the dynamics of TD events are tightly
linked to those of GW sources through the infall / inspiral connection.
The study of one bears directly on the other. One additional important
connection is due to the unfortunate fact that the electromagnetic
signature of TD events (discussed elsewhere in these proceedings)
may not be distinct enough from other types of AGN variability to
offer definitive identification. This may ultimately be achieved only
by the detection of low-freq\-uency GW counterparts to TDs.
\begin{acknowledgement}
This work was supported by ERC Starting Grant No. 202996 and DIP-BMBF
Grant No. 71--0460--0101.
\end{acknowledgement}
\bibliographystyle{epj}

\end{document}